# Innovating China's Intangible Cultural Heritage with DeepSeek + MidJourney: The Case of Yangliuqing theme Woodblock Prints

**RuiKun Yang[1], ZhongLiang Wei[1], Longdi Xian[2]**

[1]College of Business and Media, Lanzhou University of Finance and Economics
[2] Faculty of Medicine, Chinese University Of Hong Kong

First and Corresponding Author (Y13893277440@163.com).

**ABSTRACT** Yangliuqing woodblock prints, a cornerstone of China's intangible cultural heritage, are celebrated for their intricate designs and vibrant colors. However, preserving these traditional art forms while fostering innovation presents significant challenges. This study explores the DeepSeek + MidJourney approach to generating creative, themed Yangliuqing woodblock prints focused on the fight against COVID-19 and depicting joyous winners. Using Fréchet Inception Distance (FID) scores for evaluation, the method that combined DeepSeek-generated thematic prompts, MidJourney-generated thematic images, original Yangliuqing prints, and DeepSeek-generated key prompts in MidJourney-generated outputs achieved the lowest mean FID score (150.2) with minimal variability ($\sigma = 4.9$). Additionally, feedback from 62 participants, collected via questionnaires, confirmed that this hybrid approach produced the most representative results. Moreover, the questionnaire data revealed that participants demonstrated the highest willingness to promote traditional culture and the strongest interest in consuming the AI-generated images produced through this method. These findings underscore the effectiveness of an innovative approach that seamlessly blends traditional artistic elements with modern AI-driven creativity, ensuring both cultural preservation and contemporary relevance.

**Keywords:** Intangible cultural heritage; Deepseek; MidJourney; Yangliuqing theme Woodblock Prints;

## I. INTRODUCTION

Yangliuqing woodblock prints, esteemed as a significant facet of China's intangible cultural heritage (Qian, 2023), are renowned for their intricate textures, vibrant colors, and centuries-old craftsmanship (Liu, 2012). Originating during the Ming Dynasty (1368–1644) (Zhang, B., & Romainoor, N. H. 2023), these prints uniquely blend woodblock printing with hand-painting techniques (Dong Shu et al.,2024; Nag, D.,2024), reflecting traditional Chinese aesthetics and societal values (Wang, 2022). In today's rapidly modernizing society, Yangliuqing woodblock prints confront substantial challenges. Traditional production processes are intricate and time-consuming, involving meticulous handcrafting that is increasingly rare (Wang, X., & Aoki, N., 2017; Aoki, N. ,2021). This complexity hinders the prints' appeal to contemporary, digitally oriented audiences who favor more accessible art forms (Chen, C., & Wang, H., 2024; Ai, Q et al.,2024; Tsatsanashvili, A.,2024). Consequently, balancing the preservation of authenticity with the need for creative innovation has become imperative to revitalize these prints, ensuring they remain culturally relevant and economically viable (Pawar, H.,2025; Sullivan, A. M.,2015).

To address the challenges of preserving traditional art forms while introducing modern creativity, our study explores a hybrid digital methodology that integrates advanced artificial intelligence tools with traditional art references (Wang, T., & Wu, D.,2024; Gîrbacia, F.,2024). Specifically, we utilize two technological platforms: DeepSeek (Neha, F., & Bhati, D.,2025), a large language model (LLM), and MidJourney (Neha et al.,2025), an independent research lab dedicated to expanding human imaginative capabilities. This combination aims to generate new image portfolios that honor historical aesthetics while incorporating fresh creative elements (Chiou et al.,2023). DeepSeek is renowned for its advanced prompt generation capabilities, enabling the extraction of key thematic elements from traditional Yangliuqing prints (Wu et al.,2024). Its ability to produce detailed, culturally informed prompts makes it an invaluable tool for bridging the gap between historical visual language and contemporary design needs (AlAfnan, M. A. 2025). Recent industry reports highlight DeepSeek's innovative approach, significantly contributing to breakthroughs in AI-driven image synthesis (Islam et al.,2025).

Complementing DeepSeek, MidJourney functions as our primary image generation engine. Renowned for producing visually striking and detailed outputs, MidJourney translates textual prompts into high-quality images that capture both the traditional charm and modern aesthetics required for renewed cultural expression (Anantrasirichai et al.,2025). Its proven track record in digital art creation makes it particularly suitable for synthesizing the complex visual textures of woodblock prints, ensuring that the end results are consistent with historical norms and resonate with contemporary artistic sensibilities (Bakumenko, S. 2024). By integrating DeepSeek's prompt (Xian, L,2025) generation with MidJourney's image synthesis, this methodology offers a novel approach to revitalizing traditional art forms, balancing preservation with innovation (Morgado, L.,2025).

Yangliuqing woodblock prints often depict famous stories, such as the Romance of the Three Kingdoms or the love story of Niulang and Zhinu (Qian, J. 2023). Themes play a crucial role in these prints, and in the context of the 21st century (Pearce, N.,2021), the fight against COVID-19 has become an important narrative (Devera et al.,2024). The victorious battle against the pandemic is a source of collective joy, and we have selected this theme as the central focus of this study (Tao et al.,2024).

Recognizing that no single approach can fully encapsulate the multifaceted nature of Yangliuqing prints, our research will exploration of four methods for generating image portfolios. These methods were designed to test various combinations of AI-generated prompts and traditional reference imagery:

Portfolio 1: This approach utilizes DeepSeek-generated key prompts (focused on fighting COVID-19 and portraying happy winners mixed yangliuqing woodblock prints style) paired exclusively with MidJourney-generated prints.

Portfolio 2: By incorporating original Yangliuqing prints as reference images alongside DeepSeek-generated key prompts, this method enhanced consistency.

Portfolio 3: Building on the previous techniques, this method combines DeepSeek-generated theme prompts (focused on fighting COVID-19 and portraying happy winners) with MidJourney-generated theme images, while also including original prints as references.

Portfolio 4: This method integrates DeepSeek-generated theme prompts, MidJourney-generated theme images, original Yangliuqing prints, and DeepSeek-generated key prompts to create a comprehensive approach.

By fusing DeepSeek's culturally rich prompt generation with MidJourney's powerful image synthesis (Ai, Q.,2024), we can create a new paradigm in which traditional art forms can be reimagined for the modern era. This hybrid strategy not only enhances the visual quality and thematic authenticity of the generated images but also offers a scalable model for promoting cultural heritage in various digital and commercial applications. In addition to technological innovation, our work contributes to broader discussions on cultural preservation and creative industry development. As modern consumers increasingly seek products that connect them with their cultural roots in fresh and engaging ways, our findings offer valuable insights for cultural and creative product developers. The successful integration of these digital techniques can inspire similar initiatives across other domains of intangible cultural heritage, ensuring that traditional crafts not only survive but thrive in today's dynamic digital landscape.

## II. Study Framework and Methodology

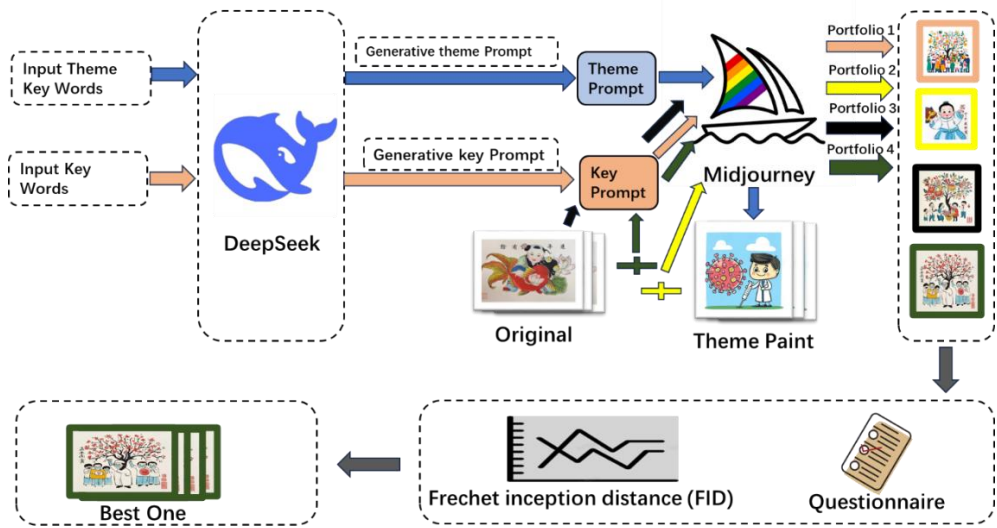

**FIGURE 1.** FrameWork of Study Work.

Figure 1 illustrated the study's workflow framework. In this study, we explored four methodologies (details shown in Methodologies) to generate Yangliuqing-style New Year woodblock prints using advanced artificial intelligence tools, specifically DeepSeek-R1 and Midjourney (details shown in overview of DeepSeek-R1 and Midjourney). Each method leveraged these AI technologies to create art that reflected the traditional aesthetics of Yangliuqing woodblock printing.

*A. Overview of DeepSeek-R1 and Midjourney*

DeepSeek-R1(DeepSeek-AI et al., 2025): DeepSeek is a Chinese AI company that has made significant advancements in artificial intelligence. Their flagship model, DeepSeek-R1, has demonstrated remarkable reasoning capabilities through reinforcement learning, achieving performance comparable to leading AI models like OpenAI's GPT series. Notably, DeepSeek-R1 was developed with a focus on reasoning without extensive supervised fine-tuning, emphasizing the model's ability to learn and adapt through reinforcement learning techniques.

Midjourney1: Midjourney is an AI-driven image generation platform that transforms textual descriptions into visual art. Accessible via Discord, users input prompts to generate images, with the platform offering various parameters to fine-tune the output, such as aspect ratio, quality, and style. Midjourney has been instrumental in enabling users to create diverse and imaginative images based on their textual descriptions.

TABLE I
The parameters for DeepSeek-R1 and Midjourney

| Tool | Parameter | Value |
|---|---|---|
| **DeepSeek-R1** | Temperature | 0.7 |
|  | Max Tokens | 150 |
|  | Top-p | 0.9 |
| **Midjourney** | Model Version | v 6.1 |
|  | Aspect Ratio (--ar) | 3:4 |
|  | Stylize (--s) | 700 |
|  | Chaos (--c) | 45 |
|  | Image Weight (--iw) | 1.5 |
|  | Seed | 3 |

---

[1] https://www.midjourney.com/

## B. Original traditional Yangliuqing prints

We collected some traditional Yangliuqing prints from (Xu et al., 2023.) as reference images, as they not only preserve the original artistic features, such as intricate line work, vibrant colors, and symbolic compositions, but also provide essential visual references to guide our AI-generated artwork. These prints serve as a benchmark for maintaining the authenticity and stylistic integrity of Yangliuqing woodblock printing in our study. An example is shown below:

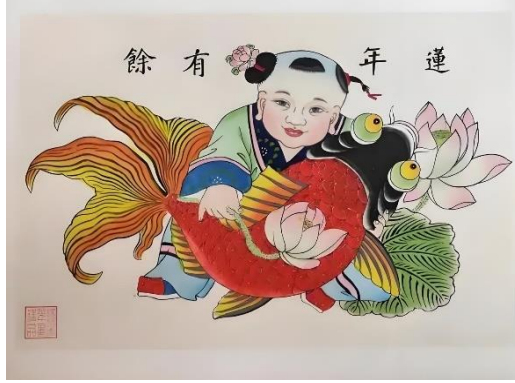

FIGURE 2. Example for Yangliuqing prints.

## C. Methodology

### 1) DIRECT PROMPT-BASED GENERATION USING MIDJOURNEY (PORTFOLIO 1)

In this approach, thematic keywords are input into DeepSeek, which then generates prompts for MidJourney based on these keywords and a predefined prompt template, details shown in table II.

TABLE II

| | Prompt Engineering Steps for Key Prompt Generation |
|---|---|
| keywords | the beautiful life of people who overcame COVID-19, now thriving in a better world |
| Prompt Template | Create a MidJourney prompt to depict{keywords}, in the style of Yangliuqing New Year woodcut prints. |
| Generated Key Prompt | Simple Yangliuqing woodcut style, white background, minimalistic figures and objects. On the left side, a group of people wearing masks, doctors and nurses holding medical tools, symbolizing the fight against COVID-19. On the right side, the same people without masks, smiling, holding flowers and celebrating, symbolizing a return to normal life. In the center, a large tree with blooming flowers, representing hope and renewal. Bright colors, clean lines, no shading, flat design. |

Midjourney then interprets this key prompt to generate corresponding images. To ensure the generated images align with the desired aesthetic, parameters such as aspect ratio, chaos level, and quality are adjusted accordingly (details shown in TABLE I).

### 2) COMBINED PROMPT AND REFERENCE IMAGE APPROACH (PORTFOLIO 2)

First, the key prompt was generated by DeepSeek (TABLE I). Simultaneously, a collection of original Yangliuqing New Year woodblock prints exemplifying the desired themes and styles was curated. These reference images were then input into MidJourney along with the prompts. MidJourney's AI model utilized both the textual key prompts and visual references from reference images to generate new prints.

### 3) THEMATIC IMAGE GENERATION FOLLOWED BY STYLE TRANSFER (PORTFOLIO 3)

In this approach, thematic keywords are input into DeepSeek to generate themed prompts.details shown in below:

TABLE III

Prompt Engineering Steps for theme Prompt Generation

| keywords | Cartoonish fighting COVID-19 virus |
|---|---|
| **Prompt Template** | Create a MidJourney prompt to depict{keywords}. |
| **Generated Theme Prompt** | A simple, colorful cartoon painting of a doctor and nurse holding giant syringes, bravely facing a cartoonish COVID-19 virus with a silly, spiky design. The background is minimal, with a cheerful blue sky and fluffy white clouds. The style is playful, with bold outlines, bright colors, and a lighthearted, optimistic tone. |

After the thematic prompt was generated, MidJourney produced theme-based paintings.

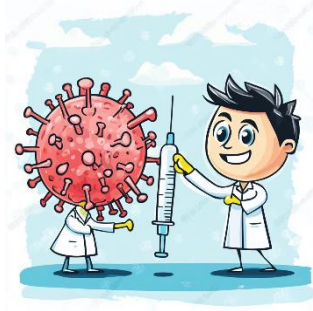

**FIGURE 3. Example of Theme Images.**

These initial AI-generated artworks captured the essence of the specified themes. Next, a selection of these generated paintings was combined with reference images of traditional Yangliuqing prints and re-input into MidJourney. This process allowed the AI to blend the stylistic elements of the original woodblock prints with the newly generated themes, resulting in refined prints that maintained the authenticity and artistic characteristics of Yangliuqing New Year woodblock printing.

4) COMBINED THEMATIC AND STYLE TRANSFER APPROACH (PORTFOLIO 4)

The thematic paintings, which combine the generated theme-based artworks with reference images of traditional Yangliuqing prints, are input into MidJourney along with the key prompt (Table II ). This process allows MidJourney to generate corresponding images that capture the desired thematic essence while adhering to the style of the reference prints. This method effectively integrates thematic image generation with style transfer, blending both textual prompts and reference images to produce artworks that reflect both the specified themes and the traditional aesthetics of Yangliuqing woodblock printing.

5) EVALUATING

Two complementary approaches were used for the evaluation of the generated images. First, we used the Frechet inception distance (FID)(Yu et al., 2021) score to quantify the similarity between the generated images and the reference Yang Liuqing prints.The FID score measures the distance between the feature distributions of the refer images and Ai-Genrated Portfolios(1 to 4), with lower FID scores indicating a higher degree of similarity.

$$FID = \| \mu_1 - \mu_2 \|^2 + Tr\left(\Sigma_1 + \Sigma_2 - 2(\Sigma_1\Sigma_2)^{\frac{1}{2}}\right), \quad (1)$$

Where $\mu_1, \mu_2$ are the means of the refer image and Ai-Genrated Portfolios, $\Sigma_1, \Sigma_2$ are the covariances of them and $Tr$ denotes the trace of a matrix.

Next, we calculated the FID scores for all AI-generated portfolios in comparison with the reference images. Finally, we obtained their mean (μ) and standard deviation (σ) for evaluating.

In addition to the computational evaluation, we conducted a user-based questionnaire (details shown in TABLE IV) to further validate the authenticity and artistic quality of the generated images. We showed participants a set of reference Yang Liuqing prints alongside the generated portfolios (Portfolio 1 to 4) and asked them to rate these portfolios in terms of cultural heritage, innovation, composition, and purchasing appeal.

By combining objective quantitative (details shown in TABLE IV) assessment with subjective human judgment, this dual approach ensured a more comprehensive assessment of the extent to which the AI-generated prints retained the essence of traditional Yangliuqing woodblock prints.

TABLE IV

## Questionnaire

| Questions | Options |
|---|---|
| What is your age? (Age) | Less than 18; 18-25; 26-30; 31-40; 41-50; 51-60; 61 or above |
| 2. What is your art-related background? (Academic Background) | Art professional practitioners; Art Enthusiast; General audience |
| 3. How much do you know about Yangliuqing woodblock prints? (Level of understanding) | Completely unaware; Somewhat aware; Very familiar |
| 4. Do you think the following four AI works have preserved the traditional colors of Yangliuqing woodblock prints? (Traditional) | Portfolio 1 (1-10); Portfolio 2 (1-10); Portfolio 3 (1-10); Portfolio 4 (1-10) |
| 5. Whether the following four works present a narrative of "drama within a painting" (fighting pneumonia and succeeding perfectly) (Theme) | Portfolio 1 (1-10); Portfolio 2 (1-10); Portfolio 3 (1-10); Portfolio 4 (1-10) |
| 6. How innovative do you think the following four AI works are? (Innovation) | Portfolio 1 (1-10); Portfolio 2 (1-10); Portfolio 3 (1-10); Portfolio 4 (1-10) |
| 7. Please rate the following works in terms of overall quality (lines, color scheme, details). (Total Quality) | Portfolio 1 (1-10); Portfolio 2 (1-10); Portfolio 3 (1-10); Portfolio 4 (1-10) |
| 8. Can these AI works pass on the cultural connotation of Yang Liu Qing prints? (Connotation) | Absolutely not; harder to pass on; average; better to pass on; perfect to pass on |
| 9. Are you willing to buy such AI-generated printmaking derivatives? (Buying) | Definitely not willing; not very willing; average; more willing; very willing |

### III. Result and Analysis

# Comparison of FID Scores

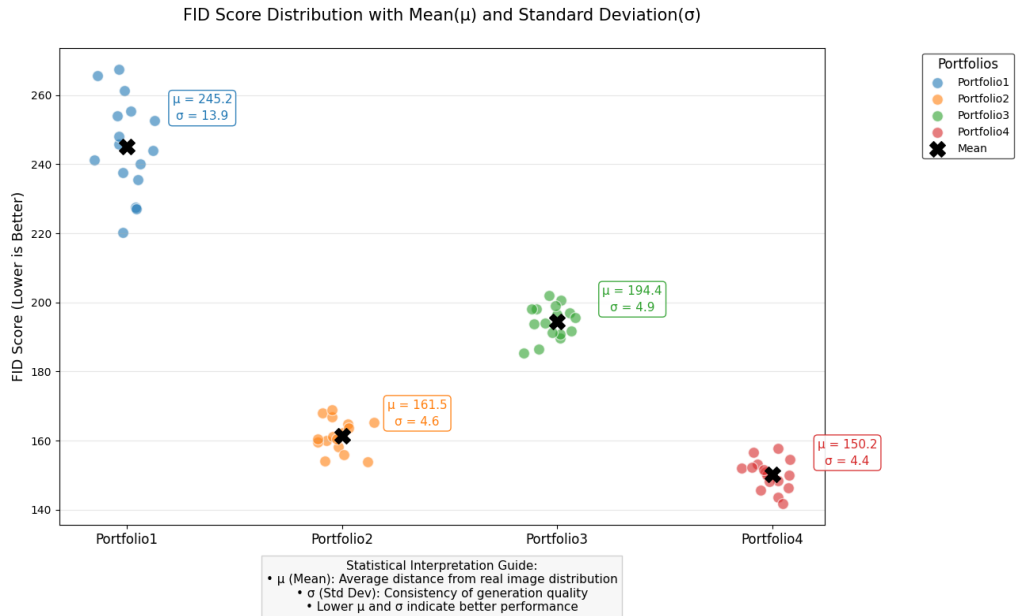

**FIGURE 4.** FID Scores Distribution Between Ai-Generated Portfolios and Refer Images.

Figure 4 illustrates the FID score distribution between AI-generated portfolios and reference images. Portfolio 1 exhibits the highest mean FID score ($\mu$ = 245.2) and the largest standard deviation ($\sigma$ = 15.3), indicating that it struggles with consistency in generating outputs that align with authentic Yangliuqing features. The substantial variability, with FID scores ranging from 220.18 to 267.44 (a range of 47.26), underscores significant deviations across generated images, suggesting that the outputs are erratic and unreliable. The high mean FID score implies that the generated images fail to accurately replicate the intricate details and aesthetic qualities of the traditional Yangliuqing style, revealing significant shortcomings in the model's understanding and execution of traditional artistic motifs. This highlights the importance of using original Yangliuqing prints as references.

Portfolio 2 demonstrates a more balanced performance, with a mean FID score of 161.5 and a relatively low standard deviation ($\sigma$ = 5.1). The FID scores are tightly clustered between 153.88 and 168.91, indicating stable performance compared to Portfolio 1. The narrow range (15.03) and low variability suggest that the model consistently generates outputs but tends to adhere closely to a standard pattern, avoiding significant variation in artistic elements. This may be due to the lack of referenced theme images, which could have introduced more creativity and diversity in the generated outputs.

Portfolio 3 falls near the borderline of acceptability with a mean FID score of 194.4 and a moderate standard deviation ($\sigma$ = 5.8). The variability in this portfolio is higher than in Portfolio 2, with FID scores showing a bimodal distribution, with peaks around 185–190 and 198–202. This suggests that while many outputs are relatively close to acceptable standards (below 200), there are occasional failures where the FID score exceeds 200, signaling the presence of significant artifacts or inaccuracies. This may be due to the absence of key prompts as references, which could have helped the model produce more consistent results.

Portfolio 4 stands out as the highest-performing model, with the lowest mean FID score ($\mu$ = 150.2) and minimal variability ($\sigma$ = 4.9). The FID scores range narrowly between 141.88 and 157.78, with a small range of 15.9. The consistency in the results is remarkable, with all FID scores falling well below 160, and the best result (141.88) approaching state-of-the-art performance.

Based on the FID analysis, Portfolio 4 emerges as the top performer, with the lowest FID score, minimal variability, and the strongest ability to preserve the traditional features of Yangliuqing woodcut art. While Portfolio 2 offers stable outputs, attributed to the use of original Yangliuqing prints and key prompts, its creative potential may be somewhat limited. In contrast, Portfolios 1 and 3 face challenges in consistency and authenticity due to the lack of key prompts or reference theme images. Therefore, the MidJourney-generated Portfolio 4, incorporating referenced key prompts, theme images, and original Yangliuqing prints, should be considered the best-performing model, offering the most reliable and authentic results.

# Questionnaire

## Participants Analysis

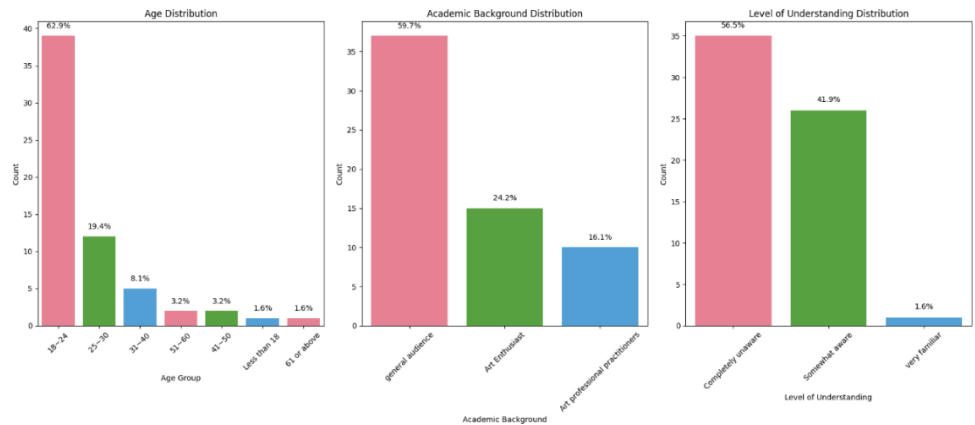

**FIGURE 5.** Questionnaire Analysis within Age, academic background and level of understanding.

Based on Figure 4, we found a total of 62 responses were collected from the questionnaire, of which the highest proportion (about 58.6%) was from the general audience aged 18-24, and the overall awareness of Yangliuqing woodblock prints was relatively low, with about 59.7% of the respondents indicating that they had no knowledge of them at all, and only one person (1.6%) being very familiar with them. Among the participants, 16.1% were art professionals and 24.2% were art enthusiasts, and their recognition was relatively high, but they still mainly had "somewhat aware". Generally speaking, the popularity of Yangliuqing woodblock prints is relatively low among young people, especially those with non-artistic backgrounds, so future promotion should focus on the general audience and strengthen cultural dissemination and education.

## Portfolio Performance Analysis

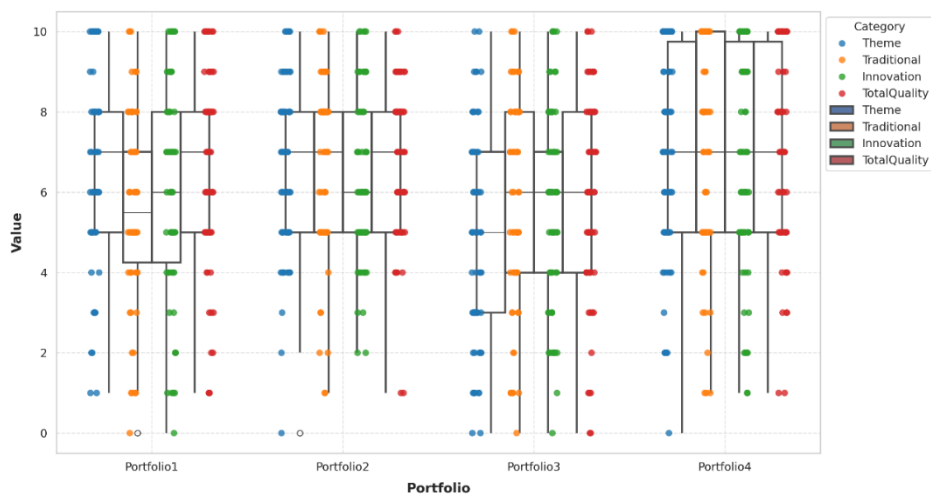

**FIGURE 6.** Portfolio Performance Distribution by Theme, Traditional, Innovation and Total Quality.

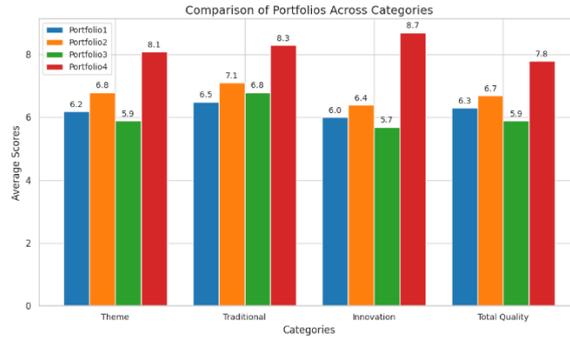

**FIGURE 7.** Comparison of Portfolios Across Categories.

After analyzing the data collected from the theme-related question (Figures 6 and 7), Portfolio 1, which utilized DeepSeek-generated key prompts and MidJourney-generated prints, achieved a moderate average score of 6.2. While some prints captured the essence of Yangliuqing themes, the high variability in scores (ranging from 1 to 10) indicates inconsistency. Additionally, the generated prints focused on thematic style rather than the original Yangliuqing artistic style (details in Figure 8).Portfolio 2, which incorporated original Yangliuqing prints as references, showed a slight improvement with an average score of 6.8. The generated images successfully integrated both the thematic focus and the authentic Yangliuqing style (details in Figure 9). However, score variability persisted, suggesting that references alone were insufficient to fully align the prints with Yangliuqing's artistic themes. Portfolio 3, which combined DeepSeek-generated theme prompts and MidJourney-generated theme images alongside original references, underperformed with an average score of 5.9. While the generated images largely adhered to the original Yangliuqing style, they failed to incorporate the intended thematic elements—particularly the absence of any COVID-19-related features, such as masks (details in Figure 10). This suggests that relying solely on images without additional prompts may have resulted in misalignment with the intended themes. Portfolio 4, which integrated DeepSeek-generated theme prompts, MidJourney-generated theme images, original references, and DeepSeek-generated key prompts, achieved the highest average score of 8.1. This approach successfully blended both thematic elements and the original Yangliuqing style (details in Figure 11). These results indicate that a multi-layered approach is the most effective for capturing the artistic themes of Yangliuqing.

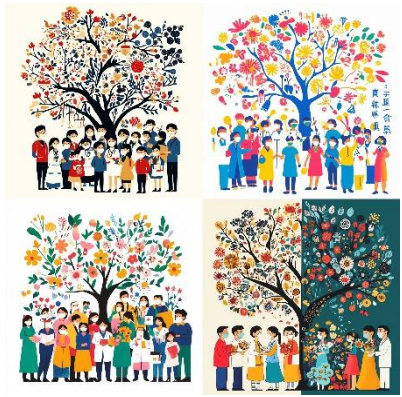
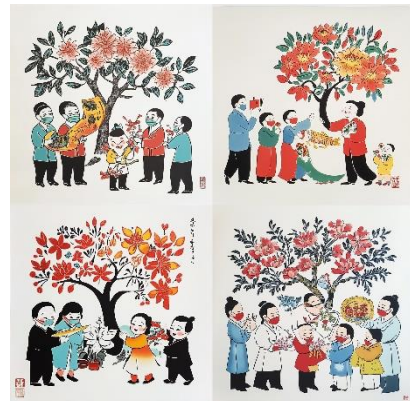

**FIGURE 8.** Example of Portfolio1.           **FIGURE 9.** Example of Portfolio2.

In the analysis of traditional style, Portfolio 1 demonstrated moderate adherence, with an average score of 6.5. However, the presence of low scores (e.g., 1 or 2) indicates significant deviations in certain prints. Additionally, the generated images lacked any original Yangliuqing style (Figure 8). Portfolio2 improved with an average score of 7.1, as the inclusion of original references helped ground the prints in traditional aesthetics. Portfolio3, despite adding theme prompts and theme images, achieved a similar average score of 6.8, suggesting that these components did not significantly enhance traditional adherence. Portfolio4 outperformed the others with an average score of 8.3, demonstrating that the combination of theme prompts, theme images, original references, and key prompts is most effective for maintaining traditional Yangliuqing style.

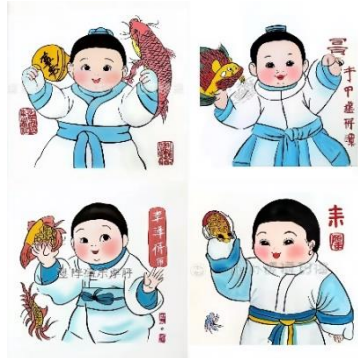
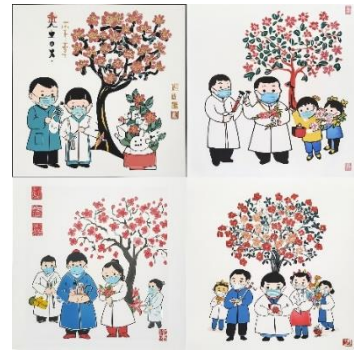

FIGURE 10. Example of Portfolio 3.      FIGURE 11. Example of Portfolio 4.

When evaluating innovation, Portfolio1 achieved a moderate average score of 6.0, but the variability in scores (ranging from 2 to 9) suggests a mix of conventional and experimental prints, with some either too traditional or too avant-garde. Portfolio2 showed a slight improvement, with an average score of 6.4 and a range from 3 to 9, as the original references allowed for a more balanced creative foundation. Portfolio3 underperformed, with an average score of 5.7 (range 2 to 8), likely due to the complexity of the method or misalignment between the components. Portfolio4 excelled, achieving an average score of 8.7, with a range of 7 to 10, demonstrating that the combination of theme prompts, images, and references facilitated greater creative freedom while still respecting traditional elements.

In terms of overall quality—considering factors like line quality, color scheme, and detail—Portfolio1 showed moderate quality, with an average score of 6.3, though the range from 2 to 9 indicates some inconsistencies. Portfolio2 improved with an average score of 6.7 (range 4 to 9), benefiting from the grounding effect of original references. Portfolio3, with an average score of 5.9 (range 3 to 8), underperformed likely due to the method's complexity. Portfolio4 again achieved the highest score of 7.8 (range 6 to 9), demonstrating that the comprehensive approach combining all elements resulted in the most polished, cohesive, and high-quality prints.

## Portfolios' Value Analysis

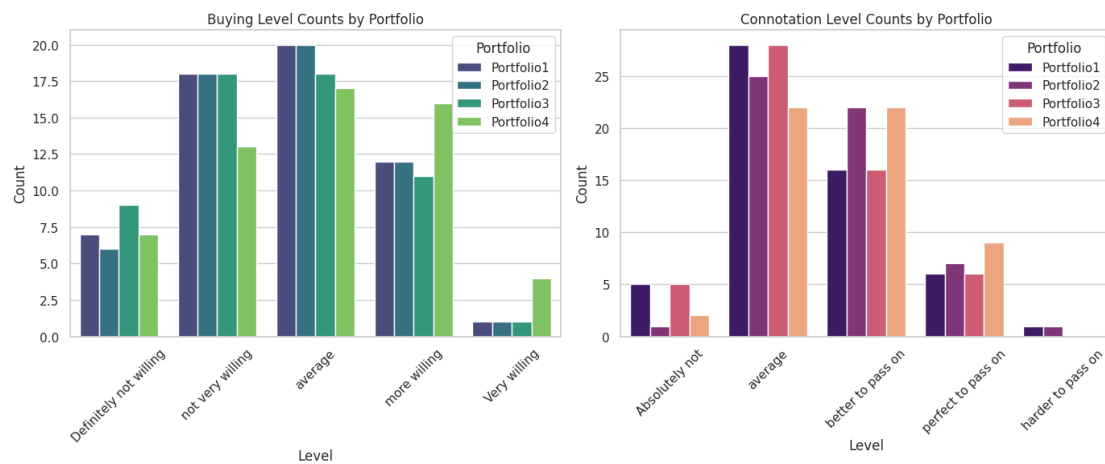

FIGURE 12. Visualization of Portfolios' Value Analysis within Buying and Connotation.

The data (details shown in Figure 12) reveals distinct patterns in participants' willingness to buy and their perception of how well each portfolio promotes traditional Yangliuqing culture. Portfolio4 stands out as the most appealing, with the highest counts in the "more willing" (16) and "Very willing" (4) categories for buying willingness, as well as the highest count in the "perfect to pass on" (9) category for connotation. This suggests that its combination of DeepSeek-generated theme prompts, original Yangliuqing prints, and MidJourney-generated prints effectively bridges modern creativity with traditional cultural elements, making it both commercially attractive and culturally resonant. In contrast, Portfolios 1 and 2 perform moderately, with most participants rating them as "average" in both buying willingness and connotation.

While they maintain a neutral appeal, they lack the strong cultural or emotional connection needed to inspire higher buying interest or deeper cultural appreciation. Portfolio3, despite its theme-based approach, shows polarization, with some participants finding it appealing while others are strongly disinterested, indicating that its thematic execution may not consistently align with cultural expectations or consumer preferences.

In terms of promoting traditional culture, Portfolio4 again excels, as its higher counts in the "perfect to pass on" and "more willing" categories suggest it successfully communicates the value of Yangliuqing traditions in a way that resonates with participants. This portfolio demonstrates how blending original prints with AI-generated elements can create a compelling narrative that honors tradition while embracing innovation. On the other hand, Portfolios 1 and 2 fall short in this regard, as their high counts in the "better to pass on" category indicate a lack of strong cultural or emotional engagement. Portfolio3, while attempting to integrate themes and original prints, fails to evoke a consistent sense of cultural promotion, as evidenced by its lack of responses in the "harder to pass on" category. Overall, the data underscores the importance of balancing modern creative techniques with authentic traditional elements to effectively promote cultural heritage and drive consumer interest.

## IV. Discussion

Yangliuqing woodblock prints are part of China's intangible cultural heritage, but innovation is currently being hindered by the challenges of preserving their intricate textures, colors, and traditional elements while incorporating creative variation. To address these challenges, we generated portfolios using four methods. The first method involved utilizing DeepSeek-generated key prompts combined with MidJourney-generated prints. The second method incorporated DeepSeek-generated key prompts alongside original Yangliuqing prints as references, with MidJourney-generated prints. The third method built on DeepSeek-generated theme prompts and MidJourney-generated theme images, adding mixed original Yangliuqing prints as references and MidJourney-generated prints. Finally, the fourth method combined a mix of DeepSeek-generated theme prompts, MidJourney-generated theme images, original Yangliuqing prints, DeepSeek-generated key prompts, and MidJourney-generated prints. And then evaluated four different portfolios using Fréchet Inception Distance (FID) scores and participant feedback to gauge the effectiveness of various methods for replicating and promoting the traditional art form.

Upon analyzing the FID scores, we found that Portfolio 1, with a mean FID score of 245.2 and a high standard deviation of 15.3, demonstrated substantial variability (220.18 – 267.44), reflecting poor consistency and difficulties in capturing the delicate details of Yangliuqing art. The lack of key reference images likely contributed to these inconsistencies, hindering the model's ability to accurately replicate the traditional style.Portfolio 2, which had a mean FID score of 161.5 and low variability ($\sigma = 5.1$), showed more consistent outputs with scores clustered between 153.88 and 168.91. However, despite its stability, this portfolio lacked the depth and creative variation needed to fully capture the richness of traditional Yangliuqing prints. The absence of specific reference theme images likely limited the model's ability to innovate within the traditional framework.Portfolio 3, with a mean FID score of 194.4 and moderate variability ($\sigma = 5.8$), exhibited mixed results. While some outputs were acceptable, others showed significant artifacts (scores above 200). This portfolio might benefit from the inclusion of referenced theme images to improve both consistency and creative output.Portfolio 4, emerging as the best performer, had the lowest mean FID score (150.2) and minimal variability ($\sigma = 4.9$). Its scores ranged narrowly (141.88–157.78), reflecting exceptional consistency and the best preservation of traditional Yangliuqing features. The superior performance of Further analysis of data from 62 participants, gathered through a questionnaire, revealed that Portfolio 4 consistently outperformed the other portfolios across all categories—theme, traditional style, innovation, and overall quality in terms of lines, color scheme, and details. This highlights the effectiveness of a hybrid approach combining DeepSeek-generated theme prompts, MidJourney-generated theme images, original Yangliuqing prints, and DeepSeek-generated key prompts. This method not only captured the essence of Yangliuqing themes but also preserved traditional aesthetics, fostered innovation, and achieved high overall quality. The inclusion of original references was crucial in grounding the prints in traditional style, while the addition of theme prompts and images enhanced thematic relevance and creativity. In contrast, the underperformance of Portfolio 3, which lacked any theme elements, suggests that without the integration of key prompts, misalignment and inconsistent results can occur. Therefore, key prompts are essential for achieving success.

Regarding participants' willingness to purchase and promote traditional culture, Portfolio 4 emerged as the most effective in driving consumer interest. It had the highest number of participants who were "more willing" (16), "very willing" (4), and deemed it "perfect to pass on" (9). The combination of DeepSeek-generated prompts, original prints, and MidJourney-generated images resonated strongly with participants, bridging modern creativity with traditional elements. In contrast, Portfolios 1 and 2 were perceived as average, lacking strong cultural or emotional engagement, while Portfolio 3 displayed polarization, indicating inconsistent appeal. This underscores the importance of integrating authentic traditional elements with innovative approaches to effectively promote cultural heritage and attract consumer interest.

## V. Conclusion and Future Work

This study investigated AI-driven methods for preserving and innovating Yangliuqing woodblock prints, a significant aspect of China's cultural heritage. Four portfolios were generated using a combination of DeepSeek and MidJourney techniques. Portfolio 4, which incorporated theme prompts, theme images, original prints, and key prompts, outperformed others in terms of Fréchet Inception Distance (FID) scores and participant feedback. It demonstrated the best balance of traditional preservation and creative innovation, achieving consistency and high quality. Portfolios 1 and 2 showed less consistency, while Portfolio 3 suffered from misalignment due to a lack of key prompts. The results underscore the importance of integrating authentic traditional elements with AI tools to maintain cultural integrity while fostering creativity.

Future research could involve refining AI models with larger datasets to improve consistency and exploring other AI techniques, such as GANs, for enhanced creativity. Expanding the study to include other traditional art forms would help assess the broader applicability of these methods.